\documentclass{pasj00}

\begin{document}
\SetRunningHead{D. Nogami et al.}{The 1996 outburst in GK Per}

\Received{2002/00/00}
\Accepted{2002/00/00}

\title{Time-Resolved Photometry of GK Per during the 1996 Outburst}

\author{Daisaku \textsc{Nogami}}
\affil{Hida Observatory, Kyoto University,
       Kamitakara, Gifu 506-1314}
\email{nogami@kwasan.kyoto-u.ac.jp}
\author{Taichi \textsc{Kato}}
\affil{Department of Astronomy, Faculty of Science, Kyoto University,
Sakyo-ku, Kyoto 606-8502}
\email{tkato@kusastro.kyoto-u.ac.jp}
\author{Hajime \textsc{Baba}}
\affil{Center for Planning and Information Systems,
  Institute of Space and Astronautical Science,\\
  Sagamihara, Kanagawa 229-8510}
\email{baba@plain.isas.ac.jp}



\maketitle

\begin{abstract}

GK Per is a unique cataclysmic variable star which has showed a nova
explosion as well as dwarf nova-type outbursts, and has the
intermediate-polar nature.  We carried out $V$-band time-resolved
photometry and $B$-band monitoring during the 1996 outburst.  This
outburst lasted about 60 d and is divided into three parts: the slow
rise branch for 35 d, the gradual decay branch with a decay rate of 20.0
d mag$^{-1}$ for $\sim$16 d, and the rapid decline branch with a rate of
5.6 d mag$^{-1}$ for $\sim$10 d.  The $B-V$ color became bluest
($B-V\sim0.18$) about 10 d before the outburst maximum, which supports
an idea that the outburst in GK Per is of the inside-out type.  The spin
pulse, $440$-s quasi periodic oscillations (QPOs), and $\sim$5,000-s
QPOs were detected in our light curve, as previously seen in X-ray and
optical observations.  In addition, we report the discovery of
$\sim$300-s periodicity, which is shorter than the spin period.

\end{abstract}

\section{Introduction}

Cataclysmic variable stars (CVs) are a group of binary stars
consisting of a white dwarf (primary star) and a late-type secondary
star (for a thorough review, \cite{war95book}).  Surface gas of the
secondary pouring out from its Roche lobe is transferred via the inner
Lagrange point to the primary.  While the gas is accreted by the white
dwarf through an accretion disk in usual CVs, the inner region of the
disk is truncated or the accretion disk is completely obstructed to be
formed in CVs containing a white dwarf with strong magnetic fields.  In
this case, the gas is forced to move along the magnetic fields and is
accreted into the magnetic poles on the white dwarf.  These magnetic CVs
are classified into two subgroups of polars having a white dwarf with a
spin period equal to the orbital period and intermediate polars (IPs)
containing a white dwarf asynchronously rotating (for a basic review,
e.g. chapter 8 and 9 in \cite{hel01book}).

CVs show two kinds of large-amplitude sudden brightenings, namely, nova
eruptions and dwarf nova-type outbursts.  The former are driven by
thermonuclear runaway reaction of the material accreted on the white
dwarf surface (\cite{sta99novareview}, and references therein).  The
amplitude is 8--15 mag.  The recurrence cycle is believed to be
typically 10$^4$ yr or more, but this cycle highly depends on the
primary mass, the mass transfer rate, the chemical abundance, and so on.
For example, the shortest recurrence period of nova eruption in U Sco is
8 yr, with recorded outbursts in 1863, 1906, 1936, 1979, 1987 and 1999.
On the contrary, the thermal instability in the accretion disk is
generally accepted as the driving mechanism for the dwarf nova-type
outburst (for a review of the disk instability model,
e.g. \cite{osa96review}).  The amplitude is 2--8 mag, and the recurrence
cycle is several days--a few years.

GK Per is the only star which has the IP nature and has shown a
nova eruption and dwarf nova-type outbursts.  On 1996 February 26, a new
outburst was caught in this enigmatic CV by J. McKenna (vsnet-obs 2211).
Following his notice, we started $V$-band time-resolved photometry and
accompanying $B$-band monitoring.  The results are reported in this
paper.  We briefly summarize the history of the GK Per study in the
literature in section 2.  Section 3 and 4 are devoted to describe the
observation and the result.  Discussion is made in section 4, and we
summarize the paper in section 5.

\section{History of the GK Per study}

GK Per is a quite unique star and has cast challenges to researchers
in various fields.  As this star has a long research history, we
here summarize a few decades.

\begin{table*}
\caption{Log of the observations.}\label{tab:log}
\begin{center}
\begin{tabular}{ccrcrcrcccc}
\hline\hline
\multicolumn{5}{c}{Date (UT)} & Orbital & N & Exposure & Band &
Magnitude$^\dagger$ & $B-V$ \\
     &          &      & &     & phase$^*$ &  & time (s) & &  \\ 
\hline
1996 & February & 26.406 & -- & 26.485 & 0.8739--0.9131 &  68 &  30 & V & 12.37 \\
     &          & 27.495 & -- & 27.549 & 0.4188--0.4462 & 223 &  15 & V & 12.35 \\
     &          &   .487 & -- &   .493 &                &  10 &  30 & B & 12.80 & 0.45 \\
     &          & 28.402 & -- & 28.578 & 0.8732--0.9611 & 632 &  15 & V & 12.11 \\
     & March    &  2.417 & -- &  2.457 & 0.3831--0.4032 &  57 &  15 & V & 11.72 \\
     &          &   .429 & -- &   .434 &                &  10 &  30 & B & 12.18 & 0.46 \\
     &          &  3.405 & -- &  3.567 & 0.8775--0.9590 & 895 &  10 & V & 11.79\\
     &          &   .399 & -- &   .404 &                &  10 &  30 & B & 12.11 & 0.32 \\
     &          &  4.433 & -- &  4.437 & 0.3925--0.3944 &   9 &  15 & V & 11.53 \\
     &          &   .425 & -- &   .430 &                &   2 &  60 & B & 11.87 & 0.34 \\
     &          &  5.402 & -- &  5.538 & 0.8779--0.9457 & 689 &  10 & V & 11.69 \\
     &          &   .396 & -- &   .402 &                &  11 &  20 & B & 12.06 & 0.37 \\
     &          &  6.393 & -- &  6.553 & 0.3741--0.4543 & 880 &  10 & V & 11.69 \\
     &          &   .397 & -- &   .402 &                &  10 &  30 & B & 12.08 & 0.39 \\
     &          & 10.403 & -- & 10.491 & 0.3821--0.4262 & 306 &  10 & V & 11.15 \\
     &          &   .399 & -- &   .402 &                &   5 &  30 & B & 11.36 & 0.21 \\
     &          & 11.398 & -- & 11.399 & 0.8803--0.8808 &   5 &  10 & V & 11.36 \\
     &          &   .399 & -- &   .401 &                &   3 &  30 & B & 11.70 & 0.34 \\
     &          & 12.395 & -- & 12.515 & 0.3795--0.4397 & 231 &  15 & V & 11.14 \\
     &          &   .397 & -- &   .400 &                &   4 &  30 & B & 11.47 & 0.33 \\
     &          & 13.467 & -- & 13.540 & 0.9164--0.9530 & 400 &  10 & V & 11.10 \\
     &          &   .468 & -- &   .470 &                &   3 &  30 & B & 11.28 & 0.18 \\
     &          & 18.399 & -- & 18.446 & 0.3859--0.4095 & 282 &  10 & V & 10.89 \\
     &          &   .403 & -- &   .404 &                &   3 &  20 & B & 11.09 & 0.20 \\
     &          & 19.429 & -- & 19.460 & 0.9018--0.9175 &  54 &  15 & V & 10.81 \\
     &          & 20.410 & -- & 20.477 & 0.3932--0.4270 & 416 &  10 & V & 10.71 \\
     &          &   .405 & -- &   .410 &                &  10 &  20 & B & 10.89 & 0.18 \\
     &          & 23.418 & -- & 23.434 & 0.8994--0.9077 &  75 &  10 & V & 10.46 \\
     &          & 26.409 & -- & 26.431 & 0.3971--0.4082 & 143 &  10 & V & 10.26 \\
     &          &   .406 & -- &   .408 &                &   5 &  20 & B & 10.60 & 0.34 \\
     & April    &  3.412 & -- &  3.437 & 0.4046--0.4175 & 109 &  10 & V & 10.64 \\
     &          &   .413 & -- &   .418 &                &  10 &  20 & B & 10.92 & 0.28 \\
     &          &  5.422 & -- &  5.477 & 0.4114--0.4389 & 349 &  10 & V & 10.71 \\
     &          &   .418 & -- &   .421 &                &   5 &  30 & B & 11.11 & 0.40 \\
\hline
\end{tabular}
\end{center}
$^*$ Calculated using the ephemeris given by \citet{mor02gkper}:
$T_{\rm 0}(HJD) = 2450022.3465 + 1.9968 \times E$.

$^\dagger$ The comparison star is HD 21588 ($V$ = 9.069, $B-V$=0.210).
\end{table*}

\begin{figure}
  \begin{center}
    \FigureFile(84mm,115mm){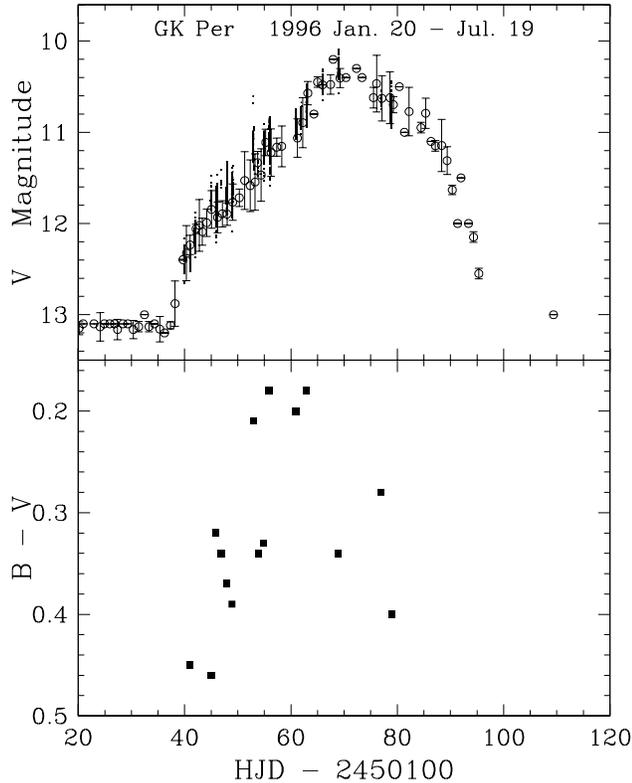}
  \end{center}
  \caption{(upper panel) The long term light curve of the 1996
    outburst of GK Per.  The open circles and the vertical bars
    represent the daily averaged visual magnitudes reported to VSNET
    and the standard deviation.  The large values of the standard
    deviation mean large-amplitude modulations of GK Per (see text).
    The small dots are our observations.  (lower panel) The $B-V$
    color.  GK Per became bluer during the outburst, and got to the peak
    about 10 days before the outburst maximum.
  }
  \label{fig:longlc}
\end{figure}

GK Per was originally discovered as a nova in 1901 (see
\cite{hal1901gkper}; \cite{pic1901gkper}).  In the decline phase of
the nova eruption, this nova showed peculiar oscillations of $\sim$1
mag with a typical cycle of several days.  The nature of these
oscillations is still unknown \citep{bia92novaoscillation}.  The
distance was derived to be 460 pc by study of the gaseous shell
ejected by the nova outburst \citep{war76shell}, although
\citet{due81novatype} quoted 525 pc. \citet{kra64oldnova} first found
that GK Per is a double-line spectroscopic binary with the secondary
star of K2 IV$p$, showing spectra having a blue continuum with the
absorption lines of the secondary star and emission lines of H, He
\textsc{i}, He \textsc{ii}, [O \textsc{ii}], and so on.  They deduced
the orbital eccentricity and the orbital period to be 0.4 and 1.904 d,
respectively.  These high eccentricity and the orbital period were
supported by \citet{bia81gkper} and \citet{bia81gkperibvs}.
\citet{cra86gkperorbit}, however, showed that a circular orbit fitted
the radial velocity variation and measured the orbital period of
1.996803 ($\pm$0.000007) d, the primary mass ($M_{\rm 1}$) of 0.9 \MO, the
secondary mass ($M_{\rm 2}$) of 0.25 \MO, and the inclination $\le73\arcdeg$.
This long orbital period has been confirmed by \citet{rei94gkperspin}
and \citet{mor02gkper} who gave a new ephemeris used in this paper.  The
spectral type of the secondary has been determined to be K0 V-III
\citep{gar94gkperspinspectroscopy}, K2V-K3V \citep{rei94gkperspin}, or
K1 IV \citep{mor02gkper}.  \citet{mor02gkper} constrained the lower
limits on the mass: $M_{\rm 1}$ $\ge0.87\pm0.24$ \MO and $M_{\rm 2}$
$\ge0.48\pm0.32$\MO.  $M_{\rm 1}$ was also estimated to be 0.36--0.86\MO,
using an X-ray observation by the ASCA satellite
\citep{ezu99MCVASCA}, although \citet{cro98MCVWDmassGINGA} tried to
estimate $M_{\rm 1}$ using Ginga data obtained in 1987, but failed.

Very Large Array (VLA) observations of GK Per carried out by
\citet{rey84gkperradio} revealed non-thermal radio emission with a
partial shell morphology, which coincided with the brightest portion
of the optical shell.  Extended emission around GK Per detected in the
far infrared IRAS imaging was interpreted to be an ancient planetary
nebula ejected from the central binary system \citep{bod87gkper}.
This was supported by radio, far-IR, and optical observations carried
out by \citet{sea89gkper}, and the 230 GHz observation in the
$J=2\rightarrow1$ line of $^{12}$CO also supports the suggestion that
GK Per is surrounded by a {\it fossil} planetary nebula
\citep{sco94gkper}.  \citet{hes89gkperdust}, however, cast doubt on
this interpretation by a large-scale survey of the $^{12}$CO dust
emission around GK Per.  The IUE ultraviolet observation made by
\citet{eva92rrpicgkperIUE} gave some evidence of abundance gradients
in the nebula.  Based on CCD imaging, \citet{anu93gkperremnant}
suggested that the nova shell is clumpy and asymmetric, and the
expansion rate is 0.31($\pm$0.07) arcsec yr$^{-1}$.
\citet{dou96gkperdust} detected, from IRAS 60 $\mu$m and 100 $\mu$m
observations, a far IR emission region extended up to 17 arcmin, which
is suggestive of a bipolar outflow.

\citet{kin79gkperXray} identified GK Per with the X-ray source
A0327+43.  This star is a very hard X-ray emitter among classical
novae \citep{bec81novaXray}.  \citet{cor81CVEinstein} noted that there
was some evidence for variability on a 50-s time scale in the Einstein
X-ray light curve.  X-ray variabilities on timescales of 100 s and hours
were also suggested by \citet{cor84CVEinstein}.  \citet{wat85gkperspin}
finally detected strong, coherent, hard X-ray ($>$2 keV) modulations
of a period of 351 s, using EXOSAT during an optical outburst in 1983,
which established the IP classification of GK Per, interpreting
the 351-s period as the spin period ($P_{\rm spin}$) of the white dwarf.  They
also found that the hard X-ray emission varies aperiodically on a
timescale of approximately 3000 s.  The 351-s pulse was observed in
quiescence with EXOSAT \citep{nor88gkperqui} and Ginga
\citep{ish92gkperGINGA}.  The pulse profile in quiescence, however, is
in a double-peaked shape, different from that in
outburst.  Ultra-high-energy radiation ($\ge$10 TeV) was not measured
from GK Per \citep{ale91ultrahighenergy}.  \citet{hel94gkper}
re-analyzed the EXOSAT data previously published by
\citet{wat85gkperspin} and found quasi periodic oscillations (QPOs) of
a timescale of $\sim$5000 s.  They proposed the disk-overflow
accretion model for these long-period QPOs.  The 351-s periodic
modulations and aperiodic intensity variations with a typical timescale
of several thousands of seconds were detected again during the 1996
outburst by ASCA \citep{ish96gkperASCAiauc}.

\begin{figure*}
  \begin{center}
    \FigureFile(168mm,230mm){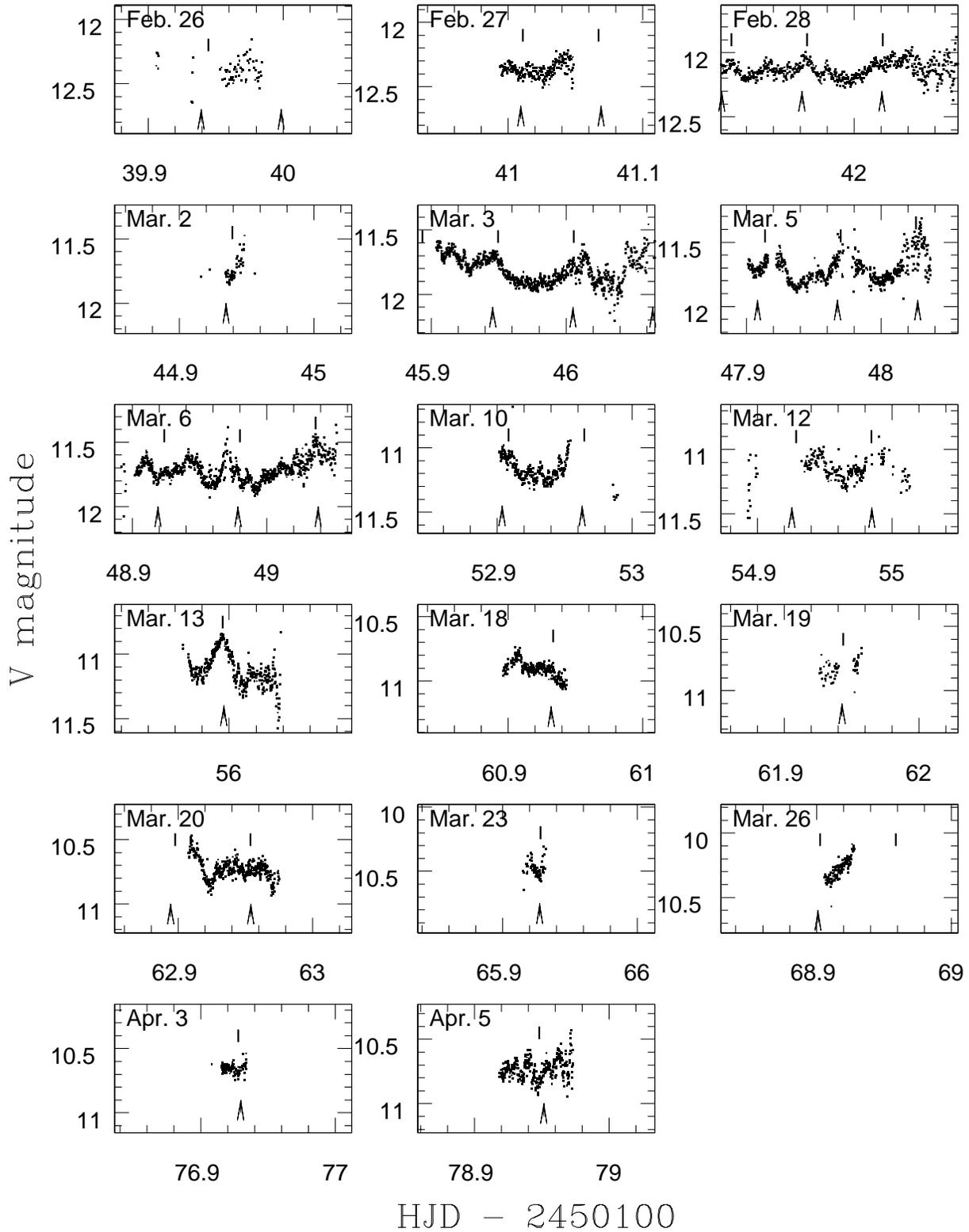}
  \end{center}
  \caption{Daily light curves.  Various-amplitude modulations, such as
    QPOs and flickerings, with various timescales were caught.  The
    tick marks and the upper arrows point at the calculated maxima of
    the 0.05612-d and the 0.05947-d oscillations (see text).
  }
  \label{fig:shortlc}
\end{figure*}

Regarding optical observations of short term variability,
\citet{rob77oldnova} reported an initial result that, among ten old
novae including GK Per, coherent oscillations were not detected except
in DQ her.  QPOs with a timescale of $\sim$380 s were subsequently
found in quiescence
\citep{pat81DNOhtcas}\footnote{\citet{pat81DNOhtcas} listed the QPO
in a table with reference to Patterson, J., 1981, ApJ, in press.  This
paper, however, have not yet been published to the authors best
knowledge.}.  U band time-resolved photometry by \citet{maz85gkper}
gave evidence of $\sim$400-s periodicity in two nights during the 1983
outburst, $\sim$360-s one during the decline phase near quiescence,
and 0.8-h one which might be an optical counterpart of the
$\sim$3000-s pulses in X-ray \citep{wat85gkperspin}.  They suggested
relationship of a beat phenomenon among $P_{\rm spin}$, $\sim$400 s, and 0.8
h.  \citet{pat91gkper} reported the discovery of the spin periodicity
in U band and QPOs of $\sim$355 s in quiescence.  During the 1986
outburst, U-band photometry performed by \citet{pez96gkper} detected
periodicity of 40 s, $\sim$370 s, 20--80 min.  In the case of
spectroscopic observations, \citet{rei94gkperspin} gave first firm
evidence for the variation of the V/R ratio of Balmer and He \textsc{i,
ii} lines with periods of $P_{\rm spin}$, 430--500 s, and 2,000--7,000 s, and
the equivalent width (EW) of these lines modulates with a period of
2,000--7,000 s (see also \cite{hut86aopscgkperpulsespec}).
\citet{gar94gkperspinspectroscopy} also found $P_{\rm spin}$ modulations of the
V/R ratio and EW of Balmer lines.  \citet{mor96gkperQPOmapping} drew
power maps of frequency versus line velocity of H$\beta$ and He
\textsc{ii} 4686 from spectra acquired during the 1996 outburst, and
detected $\sim$5,000-s QPOs.  Re-analysis of the same data by
\citet{mor99gkperQPO} showed that H$\beta$ and He \textsc{ii} have a
blue-shift bias for the 4,000--6,000-s QPOs in the power spectra of
line flux in discrete velocity bins, which supports the disk-curtain
beat mechanism rather than the disk-overflow accretion model proposed
referred to above.

{\it Dwarf nova}-type outbursts have been seen in GK Per every few
years since about 60 years after the nova explosion.  The quiescence
magnitude is around $m_V$=13.2, but has been modulating with a cycle
of a couple of decades.  This may be due to the solar-type activity in
the secondary star (\cite{ak01CVcycle}; \cite{bia90CVcycle};
\cite{war88CVcycle}).  These outbursts are characterized by peculiar
features, a long recurrence cycle of a few years, a long duration of a
few to several tens of days, a small amplitude up to 3 mag, a
double-peaked structure in ultraviolet, the rise time equal to, or
even longer than the decline time, and so on (\cite{sab83gkper};
\cite{szk85gkper}; \cite{bia86gkper}; \cite{sim02gkper}; and
references therein).  \citet{szk85gkper} obtained optical spectra
during an outburst (see also \cite{bia81gkperspec};
\cite{bia82gkper}).  He \textsc{ii} 4686 became stronger than in
quiescence, and the C \textsc{iii}/N \textsc{iii} Bowen blend
appeared.  The full widths at zero intensity (FWZI) of the emission
lines were relatively small, $\sim$40 \AA, almost equal to that in
quiescence, which implied that the accretion disk did not extend down to
the white dwarf surface either in quiescence or in outburst.  IUE
ultraviolet spectra were acquired during the rise, at the maximum,
and during the decline of the 1981 outburst \citep{wu89gkper}.  These
spectra were much redder than observed in usual dwarf novae in
outburst and in nova-likes, but \citet{wu89gkper} noted that this is
qualitatively explained by a model of the accretion disk where the
inner region is truncated by the magnetic field.  Also in quiescence,
black body fitting to an IUE spectra indicated that the temperature
was not more than 12,000 K, much lower than that of objects in this
class \citep{ros82hrdelgkperrrpicrsophUV}.  \citet{bia83gkper}
mentioned that the spectral energy distribution from X-ray to infrared
of GK Per in quiescence is better explained by a model of a disk where
magnetic fields control the temperature distribution, than another
model of a standard disk with the inner region truncated.  X-ray
started brightening $\sim$30 d earlier than optical light in the 1978
outburst \citep{bia85gkper}.  Based on this fact and that the UV
outburst light curve had two peaks, one of which occurred in the onset
phase of the optical 1981 outburst, \citet{bia86gkper} proposed that the
outbursts in GK Per start at the inner edge of the accretion disk.

\begin{figure}
  \begin{center}
    \FigureFile(84mm,115mm){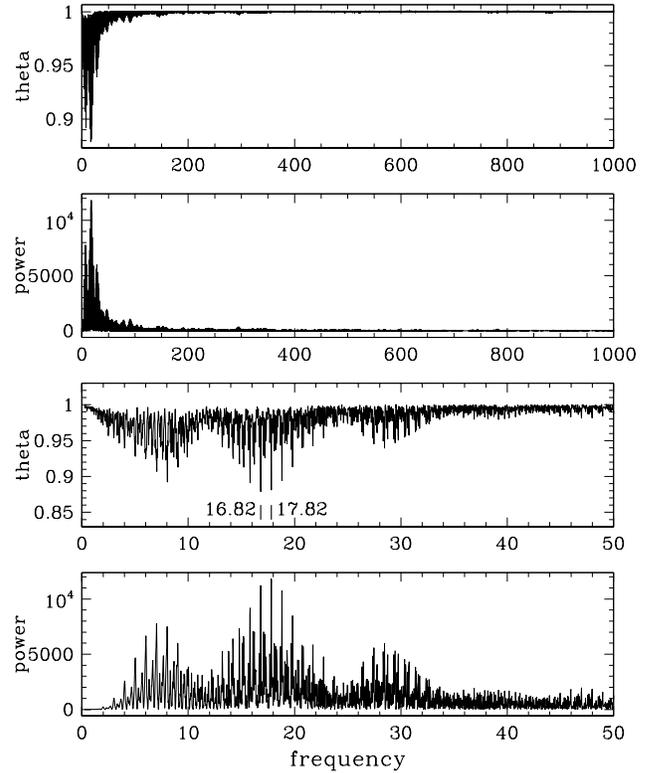}
  \end{center}
  \caption{Theta diagrams and power spectra obtained using the
    combined data of February 27 and 28, March 3, 5, 6, 10, 12, 13,
    18, and 20, and April 5.  (upper 2 panels)  Periodgrams covering
    0.25--1000 cycle d$^{-1}$.  There is no prominent peak in the
    range of $>$50 cycle d$^{-1}$.  (lower 2 panel) Same as the upper
    panels, but those enlarged below 50 cycle d$^{-1}$.  Two peaks of
    0.05947(1) d (16.62 cycle d$^{-1}$) and 0.05947(3) d (17.62 cycle
    d$^{-1}$) have almost same significance.
  }
  \label{fig:theta}
\end{figure}

A model to reproduce the peculiar outburst light curve was first
proposed by \citet{can86gkper}.  They suggested, to explain the
long outburst recurrence cycle and the long outburst duration, that
a low viscosity parameter $\alpha$, about 1/10 of the typical value for
dwarf novae with shorter $P_{\rm orb}$, and a very low mass transfer rate
$\dot{M}$ of 10$^{-9}$ \MO yr$^{-1}$ were required.  This condition
makes an eruption due to the thermal instability occur at the inner edge
of the accretion disk and the heating wave propagate to the outer disk.
Such an outburst is called {\it inside-out} outburst.  Although
\citet{ang89DNoutburstmagnetic} took into account the inner truncation
of the accretion disk due to the magnetic fields on the white dwarf, a
low $\alpha$ was still required to produce the long duration of
outburst.  \citet{kim92gkper} (see also \cite{kim92gkpererratum})
proposed another model of the {\it inside-out} outburst requiring a
larger inner radius ($\sim$2.5--4.5 $\times$ 10$^{10}$ cm) of the disk
and a large mass transfer rate ($\sim10^{18}$ g s$^{-1}$), adopting a
normal $\alpha$ value.  Introducing the stagnation phenomenon enabled
them to reproduce the double-peaked shape in UV observed by
\citet{bia86gkper}.  \citet{yi92gkper}, \citet{yi94IPhardX}, and
\citet{yi97gkper} developed this model to explain the spectral feature
in the hard X-ray and ultraviolet range during optical outburst and
during quiescence.  \citet{sim02gkper} revealed strong relationship
among the recurrence cycle, the outburst duration, and the outburst
maximum magnitude, and that they had became longer, longer, and
brighter, respectively, in the last five decades.  These changes and
that the final decay branch is very similar for all the outbursts may
be a result of three factors: decrease of irradiation of the disk by
the primary star (cf. \cite{sch01ADpostnova}), decrease of viscosity,
and consequently that the thermal instability starts at a different
distance from the disk center for the respective outbursts.

\begin{figure}
  \begin{center}
    \FigureFile(84mm,115mm){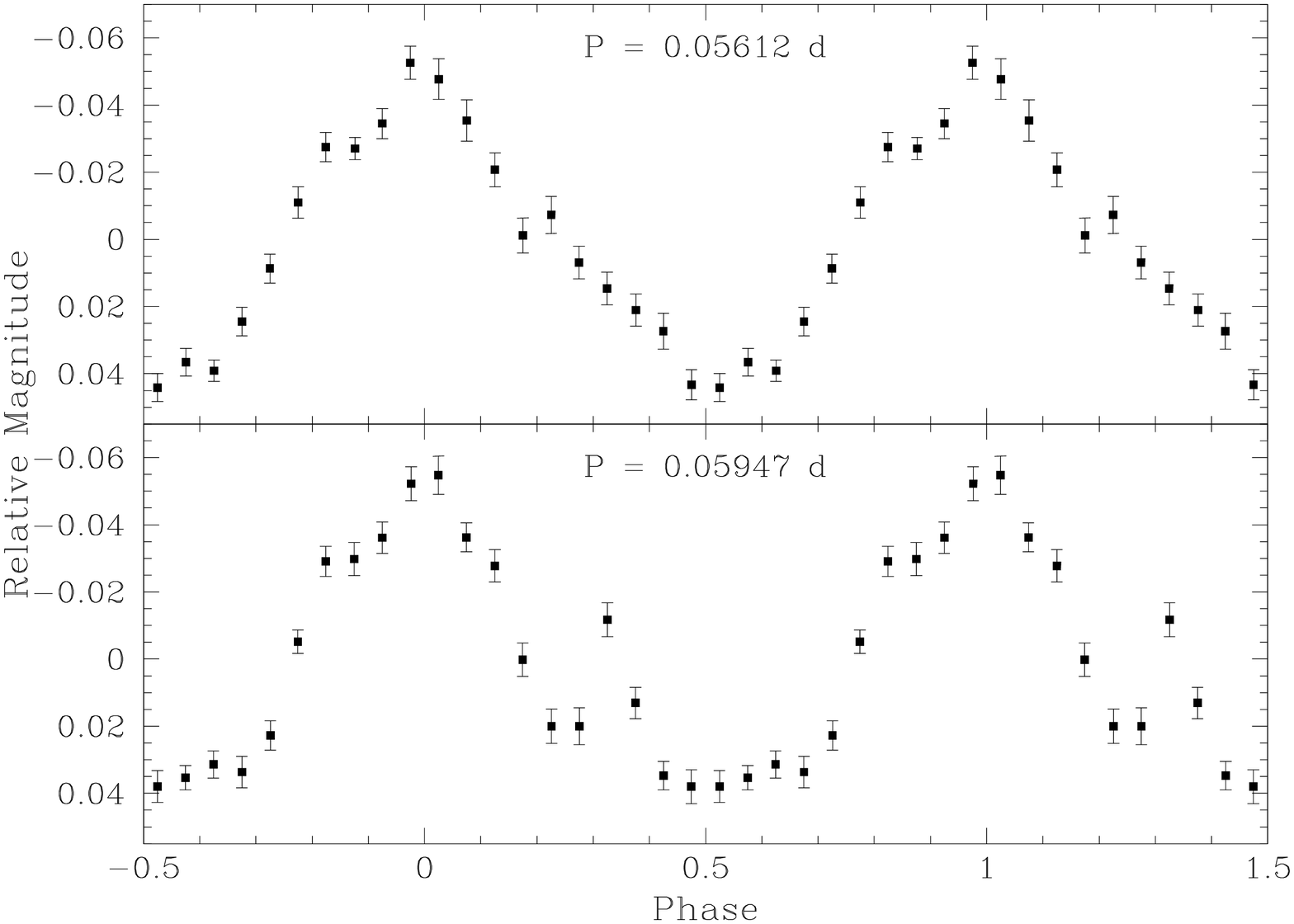}
  \end{center}
  \caption{Light curves folded with the periods of 0.05612 d and
    0.05947 d, using the data obtained on February 27 and 28, March 3,
    5, 6, 10, 12, 13, 18, and 20, and April 5.
  }
  \label{fig:qpo}
\end{figure}

\section{Observations}

We performed the observations at the Ouda Station, Kyoto University.
A 60-cm reflector (focal length=4.8 m) and a CCD camera (Thomson
TH~7882, 576 $\times$ 384 pixels) attached to the Cassegrain focus
were used (for more information of the instruments, see \cite{Ouda}).
The on-chip 2 $\times$ 2, or occasionally 3 $\times$ 3, binning mode was
selected to reduce the read-out and saving dead time.  We adopted
Johnson {\it V}- and {\it B}-band interference filters.  Table
\ref{tab:log} gives the journal of the observation.  The nominal error
for each point is typically 0.01 mag.  After standard de-biasing and
flat fielding, the frames were processed by a microcomputer-based
aperture photometry package developed by one of the authors (TK).

\begin{figure}
  \begin{center}
    \FigureFile(84mm,115mm){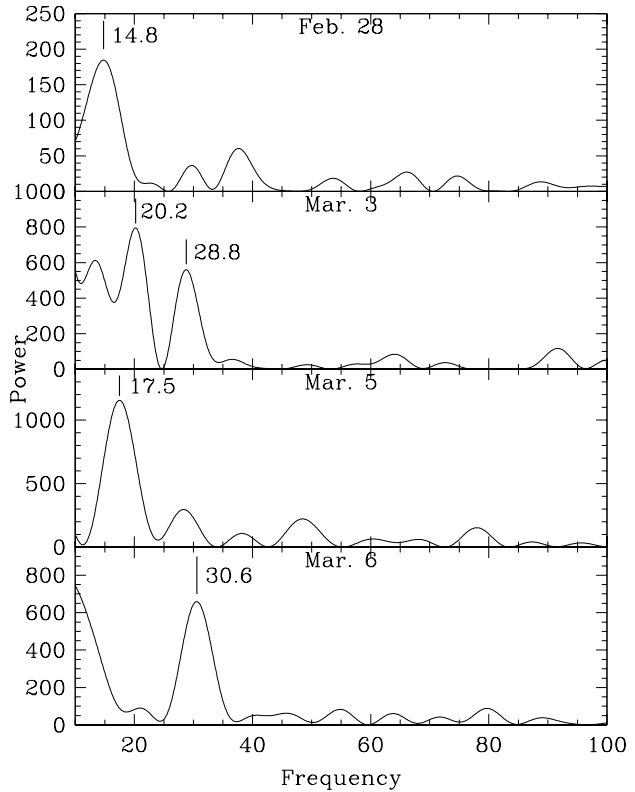}
  \end{center}
  \caption{Power spectra taken for the data on February 28, March 3, 5,
 and 6, to search for kilo-second QPOs.  The QPO periods are measured
 to be $\sim$5400 s (14.8 d$^{-1}$), $\sim$4300/$\sim$3000 s (20.2/28.8
 d$^{-1}$), $\sim$4900 s (17.5 d$^{-1}$), and 2800 s (30.6 d$^{-1}$) on
 February 28, March 3, 5, and 6, respectively.
  }
  \label{fig:lqpo}
\end{figure}

The magnitudes of the object were measured relative to a local
standard star, HD 21588 ($V$=9.069, $B-V$=0.210, spectral type A2, in
Hipparcos Input Catalogue, Version 2 \citep{HipInpCat}).  Heliocentric
corrections to observation times were applied before the following
analysis.

\begin{figure*}
  \begin{center}
    \FigureFile(168mm,200mm){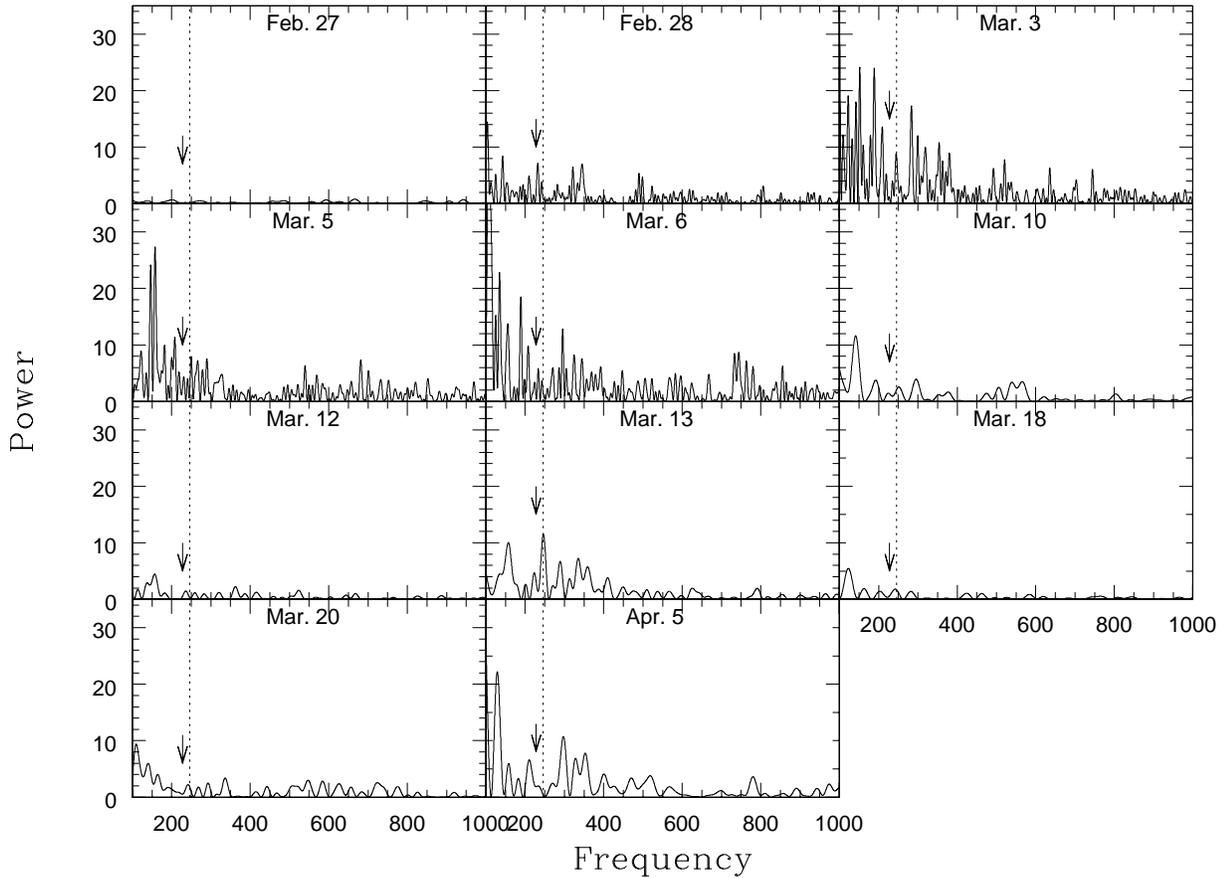}
  \end{center}
  \caption{Power spectra of daily pre-whitened data to seek short
    period modulations.  The dotted lines indicate the spin
    frequency of 246 cycle d$^{-1}$ (351 s).  The lower arrows point
    at the frequency of 229 cycle d$^{-1}$ (378 s), beat of which with
    the spin frequency gives rise to the mean QPO frequency of 5000 s.
    The spin feature is detected on March 13, and possibly on March 3.
    On March 3, and 6, QPOs with a typical period longer than the spin
    period were caught, while those with a typical period shorter than the
    spin period were detected on March 3, and 6, and April 5.
  }
  \label{fig:spin}
\end{figure*}

\begin{figure*}
  \begin{center}
    \FigureFile(168mm,200mm){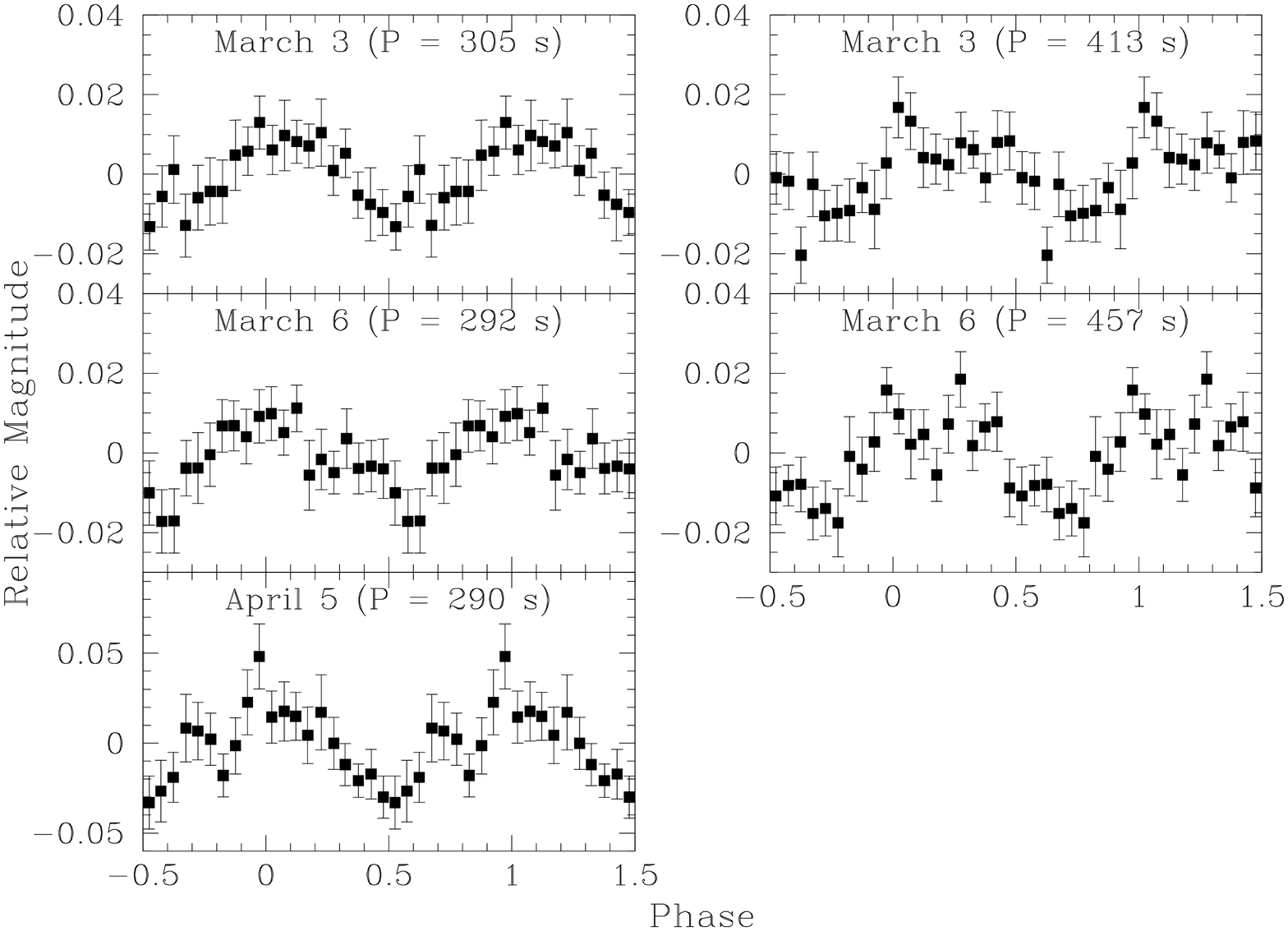}
  \end{center}
  \caption{Short QPOs detected in figure \ref{fig:spin}
  }
  \label{fig:sqpo}
\end{figure*}

\section{Results}

The long-term light curve drawn with visual observations reported to
VSNET\footnote{see $<$http://www.kusastro.kyoto-u.ac.jp/vsnet/$>$} and
our observations around this outburst is exhibited in figure
\ref{fig:longlc}.  The outburst is almost symmetric, but the rise time
is somewhat longer than the decline time.  The shape is divided into
the following three parts: 1) the slow rise with several times changes
of the rising rate to the maximum for about 35 d, 2) the subsequent
gradual decline with a rate of 20.0 d mag$^{-1}$ for about 16 d till
HJD 2450188, and 3) the following rapid decline with a rate of 5.6 d
mag$^{-1}$ for about 10 d.  The outburst duration was about 60 d and
the magnitude at the maximum is $V=10.3$.  Although the 1996 outburst
was not covered around the end because of the seasonal reason, a short
fading tail probably followed the rapid decline phase
\citep{sim02gkper}.

GK Per became dramatically bluer for the first 5 days of the outburst,
compared to its quiescence $B-V$ color of 0.9--1.0 (\cite{bia86gkper};
\cite{wu89gkper}; \cite{pat91gkper}).  The color reached its peak
($B-V\sim0.18$) about 10 d before the outburst maximum, then started
to redden.

Daily light curves are shown in figure \ref{fig:shortlc}.  We can see
modulations with amplitudes up to 0.5 mag.  Selecting data of February
27 and 28, March 3, 5, 6, 10, 12, 13, 18, and 20, and April 5 for
their relatively long coverages ($>$1 hr), we performed a period
analyses of the Phase Dispersion Minimization method \citep{PDM} and the
Fourier transfer, after subtraction of the daily linear trends, to
detect oscillations existing for a long timescale.  Figure
\ref{fig:theta} exhibits the resultant theta diagrams and power spectra.
Two peaks indicating 0.05612(1) d (= 4849 s) and 0.05947(3) d (= 5138
s), which are one-day aliases of each other, have almost same
significance.

The light curves folded with 0.05612 d and 0.05947 d, using the data
same as those for the period analyses, are drawn in the lower and
upper panels of figure \ref{fig:qpo}, respectively.  The full
amplitudes are 0.10 mag, which is much smaller than that of apparent
modulations in figure \ref{fig:shortlc}.  The ephemerides are
\begin{equation}
T_{\rm 0} (HJD) = 2450100.04336 + 0.05612 \times E,
\end{equation}
and
\begin{equation}
T_{\rm 0} (HJD) = 2450100.03474 + 0.05947 \times E.
\end{equation}
The peaks calculated by equation (1) and (2) are indicated by the tick
marks and the upper arrows in figure \ref{fig:shortlc}.  These marks do
not necessarily correspond to observed maxima, which, in conjunction
with the small amplitudes, means that the period of the modulations
varies and the 0.05612-d or 0.05947-d period is just a mean period.
Conclusively, QPOs with a typical period of $\sim$5000 s were proved to
exist during the current outburst.

For closer watch of the evolution of the kilo-second QPOs, we took power
spectra of each one-night data set covering over 3 hours.  Figure
\ref{fig:lqpo} shows that those QPOs have a period of 5400 s, 4300/3000
s, 4900 s, and 2800 s, with a error of a few hundred seconds, on
February 28, March 3, 5, 6, respectively.

We again took power spectra of the each one-night data set of
over-1-hour coverage, after pre-whitening to reveal short period
modulations more clearly.  The method of pre-whitening we adopted is to
subtract, from each point, the magnitude calculated by averaging data
within 7 minutes before and after the point.  The resultant power
spectra are shown in figure \ref{fig:spin}.
In the March 13th data, and less clearly in the March 3rd data we found,
for the first time in optical photometry obtained during outburst, clear
signs of oscillations with the spin period (350.5$\pm$2.0 s).  We
measured QPOs with a typical period of 413.0$\pm$1.6 s (209 cycle
d$^{-1}$), and 456.9$\pm$1.6 s (189 cycle d$^{-1}$) on March 3, and 6,
respectively, and marginally caught them with a period of 371.9$\pm$1.6
s (232 cycle d$^{-1}$) on February 28.  These periods are longer than
$P_{\rm spin}$.  At the same time, QPOs with a typical period of 304.7$\pm$0.9 s
(284cycle d$^{-1}$), 292.2$\pm$0.7 s (296 cycle d$^{-1}$), and
290.3$\pm$2.6 s (297 cycle d$^{-1}$) on March 3, and 6, and April 5,
were revealed.  Our observations first showed clear evidence of
existence of oscillations having a period shorter than the spin period.
The profiles of these short QPOs were depicted in figure \ref{fig:sqpo}.

The detected periods of short- and long-period QPOs are summarized in
table \ref{tab:dp}

\section{Discussion}

\subsection{short- and long-term QPOs}

As described in the previous section, we detected two periodicities
for short-term QPOs, which are $\sim$300s and $\sim$370--460s, on
February 28, March 3, 6, and April 5 during our observation since 1996
February 26.  Since the amplitude of the QPOs varies with a timescale of
hours (\cite{pat91gkper}; \cite{maz85gkper}), it is likely that QPOs had
amplitudes smaller than the detection limit on the other nights.

\begin{table*}
\caption{Detected QPO periods.}\label{tab:dp}
\begin{center}
\begin{tabular}{lrcccc}
\hline\hline
\multicolumn{2}{c}{Date} & \multicolumn{3}{c}{Short period (s)} & Long period (s)\\
  &        & $<$$P_{\rm spin}$  & $P_{\rm spin}$  & $>$$P_{\rm spin}$ & \\
\hline
Feb. & 28   &            &         & 371.9($\pm$1.6)? & $\sim$5400 \\
Mar. & 3    & 304.7($\pm$0.9) & ? & 413.0($\pm$1.6) & $\sim$4300/$\sim$3000\\
Mar. & 5    &            &          &                 & $\sim$4900 \\
Mar. & 6    & 292.2($\pm$0.7) &  & 456.9($\pm$1.6) & $\sim$2800 \\
Mar. & 13   &            & 350.5($\pm$2.0) & & $^*$ \\
Apr. & 5    & 290.3($\pm$2.6) &  & & $^*$ \\
\hline
\multicolumn{6}{l}{$^*$ The coverage is not sufficient for the
 period analysis for kilo-second QPOs.}
\end{tabular}
\end{center}
\end{table*}

\citet{mor99gkperQPO} spectroscopically observed GK Per on 1996
February 26, 27, and 28, and reported results of detailed period
analyses.  While kilo-second QPOs and spin pulsations were detected
in the V/R ratio variation of H$\beta$ and He \textsc{ii} 4686 in all
nights except for spin pulsations in He \textsc{ii} on February 26, they
also caught kilo-second QPOs in the integrated line fluxes of H$\beta$
and He \textsc{ii} and the continuum in all nights.  Spin pulsations
in the integrated line fluxes were found only on February 26.  The
typical period of the kilo-second QPOs of the continuum around H$\beta$
and He \textsc{ii} 4686 in \citet{mor99gkperQPO} was 4477 ($\pm$ 12) s
(= 0.0518 d; 19.3 cycle d$^{-1}$) on February 27 and 3692 ($\pm$ 9) s (=
0.0427 d; 23.4 cycle d$^{-1}$) on February 28.  However, spin pulsations
were not detected in our data on February 27, and 28.  Contribution of
the continuum would dilute the spin pulsations.  The 5400-s period we
detected on February 28 was far longer than 3692 s found by
\citet{mor99gkperQPO} on the same day.  This may suggest that the QPO
period depends on the wavelength.

The period of the kilo-second QPOs observed by \citet{mor99gkperQPO}
decreased from night to night in February 26-28, and they interpreted
that this might be due to shrinkage of the blobs' orbit.  The
kilo-second QPO period, however, once increased on March 5 (table
\ref{tab:dp}).  This period increase may be explained by generation of a
new blobs on the outer orbit, but note that the period rapidly decreased
to $\sim$2800 s by the next night.

Concerning short-term QPOs, we found two typical periods of $\sim$300
s and $\sim$440 s in $V$ band.  The current standard scheme of QPOs in
GK Per is the beat frequency model which was originally proposed by
\citet{lam85QPO} for QPOs in the X-ray binary Sco X-1.  For QPOs in GK
Per, \citet{wat85gkperspin} first interpreted the X-ray QPOs with this
model that short-term QPOs are originated in dense blobs orbiting near
the inner edge of the accretion disk with the Keplerian frequency and
long-term QPOs observed in X-ray and optical has the beat frequency
between the spin frequency and the blobs' frequency.  Developing this
model, \citet{mor99gkperQPO} proposed a disk-curtain beat mechanism.
The other plausible model is the disk-overflow accretion model proposed
by \citet{hel94gkper}.  This model is based on the idea that the
kilo-second period is the characteristic timescale of the radius where
an accretion stream overflowing the disk would collide back onto the
disk.

Simple calculation by the beat model yields the long-term QPO period of
1,500 s--2,400 s for both of $\sim$440-s and $\sim$300-s periodicities
we observed.  These periods, however, can not be reconciled with the
period of 2800--5400 s in our data, which might favor the disk-overflow
accretion model.  But the coverages of our data were not neccesarily
long enough for accurate determination of the kilo-second periods.  It
is important to organize a multi-longitudinal observation campaign to
simultaneously investigate periodicities of hundreds of seconds and
kilo-seconds.  Note that using a bluer band is better in this case,
since the amplitude of QPOs is larger in a bluer band (see
\cite{pat91gkper}).

Oscillations with a period of $\sim$300 s are the first discovery of
below-the-$P_{\rm spin}$ modulations except for the 40-s pulse reported by
\citet{pez96gkper} whose origin is still unknown.  If this period is
the Keplerian period of blobs, the inner edge of the accretion disk
approached the white dwarf surface over the co-rotation radius during
the 1996 outburst, while the inner radius in quiescence is required
to be far larger than the co-rotation radius \citep{kim92gkper}.  The
QPO period in fact decreased from 305 s to 290 s in March 3--April 5.
Theoretical exploration on how closely the inner radius can approach
the primary star against the magnetic torque during outburst should be
an interesting topic.  It should be also pointed out that two
periodicities were caught on the same nights: 305 s and 413 s on March 3
and 457 s and 292 s on March 6.

\begin{table*}
\caption{Properties of IPs showing outbursts.}\label{tab:ip}
\begin{center}
\begin{tabular}{ccccccc}
\hline\hline
          & $P_{\rm orb}$ & $P_{\rm spin}$ & Outburst & Outburst & Recurrence & Reference\\
          &  (d)  &  (s)  & duration(d) & amplitude (mag) & cycle (d) & \\
\hline
HT Cam    & 0.0604 & 514 & $\sim$2 & 4.5 & 150 & 1, 2 \\
EX Hya    & 0.0682 & 4022& 2--3  & 3.5 & $\sim$550$^*$ & 3, 4 \\
V1223 Sgr & 0.1402 & 794 & $\sim$0.5 & $<$0.8 &  & 5, 6 \\
DO Dra    & 0.1650 & 529 &   5   & $\sim$5 & $>$300 & 7, 8, 9, 10, 11,
          12, 13\\
TV Col    & 0.2286 & 1911& $\sim$0.5 & 1--2$^\dagger$ &  & 14, 15, 16 \\
XY Ari    & 0.2527 & 206 &   5   &     &      & 17, 18, 19 \\
GK Per    & 1.9968 & 351 & $>$60 & 3.0 & $\sim$1000 & 20, 21, 22 \\
\hline
\end{tabular}
\end{center}

Note: In some case, we derived the values of outburst duration,
outburst amplitude, and the recurrence cycle from the observations
reported to VSNET.

$^*$ In only one case, an outburst was caught just 8 days after
the previous one.

$^\dagger$ These large amplitude outbursts occur only when TV Col is
in high state \citep{aug94tvcol}.

Reference: 1 \citet{tov98htcam}, 2 \citet{ish02htcam}, 3
\citet{hel92exhya}, 4 \citet{hel00exhyaoutburst},
5 \citet{vaname87v1223sgr}, 6 \citet{jab87v1223sgr}, 7
\citet{wen83dodra}, 8 \citet{mat91dodraIR}, 9 \citet{pat93dodraXray},
10 \citet{has97dodraHST}, 11 \citet{nor99dodrav709cas}, 12
\citet{sim00dodra}, 13 \citet{szk02dodra}, 14 \citet{sch85tvcolspin},
15 \citet{hel93tvcol}, 16 \citet{aug94tvcol}, 17 \citet{tak89xyariiauc},
18 \citet{kam91xyari}, 19 \citet{hel97xyariRXTEoutburst}, 20
\citet{wat85gkperspin}, 21 \citet{mor02gkper}, 22 \citet{sim02gkper}

\end{table*}

\subsection{Outburst mechanism}

Outbursts in dwarf novae are currently thought to be caused by the disk
instability.  However, the outburst mechanism in GK Per is still an open
problem, because of the IP nature, the peculiar binary parameters such
as the quite long $P_{\rm orb}$, and the extraordinary outburst property shown
in figure \ref{fig:longlc} and summarized in table \ref{tab:ip}.  While
\citet{bia82gkper} explained the 1981 outburst by an increase in the
mass transfer rate from the late-type secondary, \citet{can86gkper},
\citet{ang89DNoutburstmagnetic}, and \citet{kim92gkper} proposed models
based on the disk instability scheme, as mentioned in section 2.

SS Cyg-type dwarf novae usually show asymmetric outbursts with a rapid
rise and a slow decay, but some stars occasionally cause rather
symmetric outbursts with slow rise, say, ``anomalous'' outbursts (for
example, \cite{can92sscyg} and references therein for the SS Cyg
case).  Theoretical investigations indicate that the former is the
outburst of the {\it outside-in} type and the anomalous outburst is
the one of the {\it inside-out} type \citep{sma84DI}.  The outbursts in
GK Per are very similar to the anomalous outburst in shape, although the
timescales of rise and decline are much longer in GK Per than in usual
dwarf novae.  Therefore, \citet{can86gkper},
\citet{ang89DNoutburstmagnetic}, and \citet{kim92gkper} attempted to
apply an idea of the {\it inside-out} outburst for explanation of the
outburst pattern of GK Per.

Our observations revealed that the $B-V$ color reached its minimum
about 10 days before the outburst maximum.  This agrees with the fact
that SS Cyg was bluest in the $B-V$ color 4.0 d before the $V$-band
maximum during the 1981 September anomalous outburst
\citep{hop84sscyganomalousoutburst}, though SS Cyg become bluest at, or
a little later than the maximum during normal outbursts
(e.g. \cite{hop80sscyg}).  The $B-V$ observations presented here support
that outbursts in GK Per are of the {\it inside-out} type.

Although \citet{can86gkper} did not take into account the inner-disk
truncation due to the magnetic fields on the white dwarf, their
calculation provided a prediction that GK Per becomes bluer in $B-V$ for
a while after the outburst maximum, which clearly contradicts with our
result.  In polishing up the model for GK Per being rooted in the disk
instability theory, it may be a key whether the model can account for
the $B-V$ variation during outburst.

\citet{kim92gkper} proposed a time-dependent disk instability model to
reproduce the recurrence time, the duration, the doubly-peaked
structure in the UV range.  This model appears to be able to explain
also the outburst shape consisting of the three parts (see e.g. figure 6
in their paper), although the radius of the inner edge must be
changed for the individual outburst in order to reproduce the
variation of the rise-branch shape.  \citet{sim02gkper} found that the
rise branch had showed large scatter while the decline rate during the
final rapid decline was common to 8 outbursts in these three decades.

We here focus on the final rapid decline.  During this part, the
cooling front from the outer region propagates onto the inner edge (see
figures in \cite{kim92gkper}).  This process is common to the outbursts
of the outside-in type and of the inside-out type, so that the decline
rate during this phase is expected to be on the Bailey's relation
(\cite{BaileyRelation}; \cite{szk84AAVSO}; \cite{war95book}).  The value
of 5.6 d mag$^{-1}$ in figure \ref{fig:longlc}, however, is much smaller
than 10.0 d mag$^{-1}$ calculated using the Bailey's relation for the
$P_{\rm orb}$ of GK Per.  \citet{sim02gkper} derived 8.0 d mag$^{-1}$ from the
recent 8 outbursts [they also derived a value of 14.5 d mag$^{-1}$ using
the data including those during the gradual fading.  However, this
treatment is not adequate because different processes are thought to
work during the gradual fading phase and during the rapid fading
phase].  Also in other IPs, DO Dra \citep{sim02gkper} and HT Cam
\citep{ish02htcam}, the decay timescale is reported to be shorter than
expected from their $P_{\rm orb}$ by the Bailey's relation.  The decay timescale
is made shorter by the inner truncation of the disk
(\cite{ang89DNoutburstmagnetic}; \cite{can94DNBHXNdecay}), and the
magnetic fields penetrating the accretion disk may also have an effect
on the timescale.

Properties of GK Per and other outbursting IPs are summarized in table
\ref{tab:ip}.  We can not see a simple correlation between any two
factors among 5 systems.  This means that it is difficult to establish a
unified scheme on the outburst mechanism in IPs.  Note that short
outbursts in V1223 Sgr and TV Col are possibly caused by the mass
transfer events, not by the disk instability (e.g. \cite{hel93tvcol}).
A modeling including the effects of magnetic fields penetrating and
truncating the accretion disk, 3D structures of the accretion disk, the
mass stream, the accretion curtain, and the accretion pole, will be
required to reproduce a variety of outbursts in IPs.

\section{Conclusion}

Our $V$-band time-resolved photometry revealed the spin pulse ($P_{\rm
spin}$ = 351 s), and QPOs with three typical timescales of 300 s, 440 s,
and 5,000 s.  The discovery of 300-s QPOs means that the inner edge of
the accretion disk shrinks over the co-rotation period during outburst.
The beat period of 440 s and 351 s is $\sim$1,700 s, which would be
against the model for the kilo-second QPOs.  Further check of
contemporaneousness of hundred-second and kilo-second QPOs is required.
The $B-V$ color became bluest ($B-V\sim0.18$) about 10 d before the
outburst maximum.  A similar color variation was observed during an
anomalous outburst in SS Cyg, and this agrees with the idea that GK Per
shows inside-out-type outbursts.  The decline rate during the rapid
decline phase is smaller in GK Per, as well as DO Dra, and HT Cam, than
expected from the $P_{\rm orb}$.  This is qualitatively explained by truncation
of the inner part of the accretion disk, but more detailed modeling
taking into account the effect of the magnetic fields, 3D structures of
the accretion disk, the mass stream, the accretion curtain, and the
accretion pole is needed to reproduce the whole outburst properties of
IPs.

\vskip 3mm

We thank the anonymous referee for the useful comments.  The authors are
very grateful to amateur observers for continuous reporting their
valuable observations to VSNET.  Thanks are also to Emi Nakaoka for her
help in searching for the literature.  This work is partly supported by
a grant-in-aid (13640239) from the Japanese Ministry of Education,
Culture, Sports, Science and Technology (TK).

\end{document}